\documentclass[twocolumn,showpacs,preprintnumbers,amsmath,amssymb,superscriptaddress]{revtex4}
\usepackage{graphicx}% Include figure files
\usepackage{dcolumn}% Align table columns on decimal point
\usepackage{bm}% bold math
\begin{document}
\title{Improved radiative corrections  and proton charge form factor from
the Rosenbluth separation technique}

\author{Florian Weissbach}
%\email{florian.weissbach@gmail.com}
\affiliation{%
GSI Helmholtzzentrum f\"ur Schwerionenforschung GmbH, 
D -- 64291 Darmstadt, Germany
}%
\affiliation{%
Departement f\"ur Physik,
Universit\"at Basel,
CH -- 4056 Basel, Switzerland}%
\author{Kai Hencken}
%\email{k.hencken@unibas.ch}
\affiliation{%
Departement f\"ur Physik,
Universit\"at Basel,
CH -- 4056 Basel, Switzerland}%
\affiliation{%
ABB Switzerland Ltd., Corporate Research, CH -- 5405 Baden-D\"attwil, 
Switzerland}%
\author{Dirk Trautmann}
%\email{dirk.trautmann@unibas.ch}
\affiliation{%
Departement f\"ur Physik,
Universit\"at Basel,
CH -- 4056 Basel, Switzerland}% 
\author{Ingo Sick}
%\email{ingo.sick@unibas.ch}
\affiliation{%
Departement f\"ur Physik,
Universit\"at Basel,
CH -- 4056 Basel, Switzerland}%

\date{\today}

\begin{abstract}
We investigate whether the apparent discrepancy between proton electric 
form factor from measurements using the Rosenbluth separation
technique and polarization transfer method is 
due to the standard approximations employed in radiative correction procedures.
Inaccuracies due to both the peaking approximation and the soft-photon 
approximation have been removed in our simulation approach. In contrast to results from
$(e,e'p)$ experiments, we find them in this case to be too small to explain the
discrepancy.
\end{abstract}


\maketitle
\section{Introduction}
Knowledge of the electromagnetic form factors of the proton and 
the neutron, $G_{\rm ep}$, $G_{\rm mp}$, $G_{\rm en}$, and $G_{\rm mn}$, 
is important for an understanding of the inner structure of the nucleon.
Until recently the proton form factors $G_{\rm ep}$ and $G_{\rm mp}$ have 
been determined from {\em e-p} cross section measurements using the 
Rosenbluth technique, {\em i.e.}~by measuring cross sections at constant 
momentum transfers 
$Q^2=-q^2$ 
at forward and backward scattering angles. 
More recently, the  polarization technique has
become available for proton form factor measurements; 
the recoil proton polarization in {\em e.g.}~$\vec{e}${\em -p} 
scattering yields the ratio $G_{\rm ep}/G_{\rm mp}$.\\

The Rosenbluth separation technique is based on the assumption that the 
interaction between electron and proton occurs via a single-photon exchange 
(Born approximation).
This assumption leads to an $e-p$ cross section from which $G_{\rm ep}$
and $G_{mp}$ can be deduced as follows \cite{rosenbluth}.
Defining the variable $\varepsilon$ as
\begin{eqnarray}
\label{rosenbluthepsilon}
\varepsilon^{-1}\equiv1+2(1+\tau)\tan^2\frac{\theta}{2} \, ,
\end{eqnarray}
where $\tau\equiv -q^2/(4M^2)$, with $M$ being the proton mass and $q$
the (four) momentum transfer,
the $e-p$ cross section in terms of the Mott cross section yields
\begin{eqnarray}
\label{epepsilon}
\left(\frac{d\sigma}{d\Omega_{\rm e}}\right)_{\rm ep}
=\left(\frac{d\sigma}{d\Omega_{\rm e}}\right)_{\rm Mott}
\frac{\tau G_{e_m}^2+\varepsilon G_{\rm ep}^2}{\varepsilon(1+\tau)} \, .
\end{eqnarray}
We then define the so-called reduced cross section by
\begin{eqnarray}
\label{defredcross}
\sigma_{\rm red}\equiv
\left(\frac{d\sigma}{d\Omega_{\rm e}}\right)_{\rm ep}
\frac{\varepsilon(1+\tau)}
{\left(\frac{d\sigma}{d\Omega_{\rm e}}\right)_{\rm Mott}}
\, .
\end{eqnarray}
Inserting cross section (\ref{epepsilon}) into
(\ref{defredcross}), the reduced cross section becomes
a linear function in $\varepsilon$ \cite{rosenbluth},
\begin{eqnarray}
\label{redcrosslinear}
\sigma_{\rm red}=
\tau G_{\rm mp}^2 + \varepsilon G_{\rm ep}^2 \, .
\end{eqnarray}
The slope of this linear function equals
$G_{\rm ep}^2$ and its intercept is $\tau G_{\rm mp}^2$.
Comparing forward and backward scattering, each set of measurements at 
constant $Q^2$ yields one data point of the form factors $G_{\rm ep}^2(Q^2)$ 
and $G_{\rm mp}^2(Q^2)$ at the chosen momentum transfer $Q^2$.\\

Cross sections depend on the proton electric and magnetic
form factors simultaneously which poses a problem at higher values of $Q^2$ 
where the respective contributions of $G_{\rm ep}$ and $G_{\rm mp}$
to the reduced cross section (\ref{redcrosslinear}) are distributed
un-evenly among the two form factors. 
At {\em e.g.}~$Q^2=5\,{\rm GeV^2}$, the electric form factor contribution 
to the reduced cross section is down to $8\%$ and it further decreases with 
increasing $Q^2$.
Hence the slope of the measured reduced cross section (\ref{redcrosslinear})
becomes extremely small and thus very sensitive to systematic errors.\\

Form factor measurements are often parameterised using the so-called dipole form factor
\begin{eqnarray}
\label{dipole}
G_{\rm d}(q^2)=
\left(\frac{1}{1-\frac{q^2}{\Lambda^2}}\right)^2 \, ,
\end{eqnarray}
where the term 'dipole' refers to the two poles of the denominator;
$\Lambda$ is a constant of the order of $1\,{\rm GeV}$.
To date all Rosenbluth measurements are approximately compatible with sca\-ling,
{\em i.e.}~all Rosenbluth experiments indicate that
\begin{eqnarray}
\label{scaling1}
G_{\rm ep}\sim G_{\rm d} \,\,\,\,
{\rm or}\,\,\,\,
G_{\rm ep}\sim G_{\rm mp}/\mu_{\rm p} \, ,
\end{eqnarray}
respectively \cite{arrington,christy,andivahis}, 
where $\mu_{\rm p}$ is the proton magnetic moment.\\

In contrast to Rosenbluth measurements, polarization transfer experiments 
use polarized electron beams or polarized proton targets.
In nuclear physics nomenclature they are denoted {\em e.g.}~as 
$(\vec{e},e'\vec{p})$ reactions.
Polarization transfer experiments with maximum values of $Q^2$ large 
enough to exhibit a discrepancy with results from Rosenbluth measurements 
were carried out at {\sc tjnaf} in 1998 \cite{recpol}.
They covered momentum transfers from $Q^2=0.5\,{\rm GeV}^2$ to 
$3.5\,{\rm GeV}^2$.
By swapping spectrometers and using a calorimeter for electron detection, 
higher momentum transfers of up to 
$5.6\,{\rm GeV}^2$ became accessible later \cite{recpol2}.\\

\begin{figure}%[t]
\centering
\includegraphics[width=8cm]{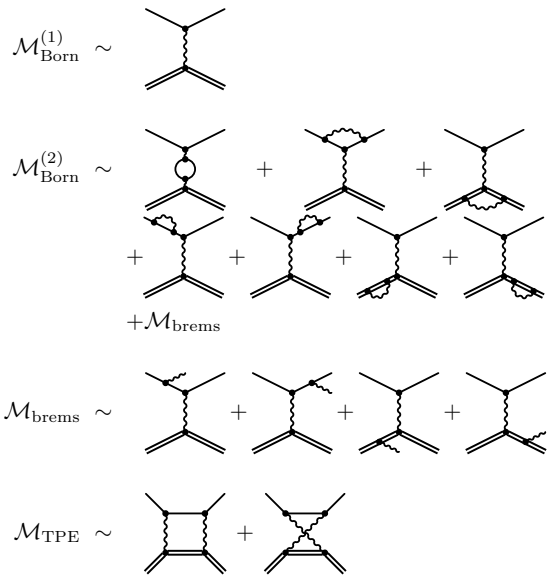}
\caption{\label{fig1} Feynman diagrams beyond the leading order.
${\cal M}_{\rm Born}^{(1)}$ and ${\cal M}_{\rm Born}^{(2)}$ constitute 
the Born approximation. The latter amplitude includes the internal radiative
corrections from vacuum polarization, vertex corrections, and 
self-energy diagrams; and the external radiative corrections
referred to as brems\-strah\-lung from ${\cal M}_{\rm brems}$.
Two-photon exchange {\sc (tpe)} contributions are not included in the
Born approximation.
}
\end{figure}
\begin{figure}[t]
\centering
\includegraphics[width=9cm]{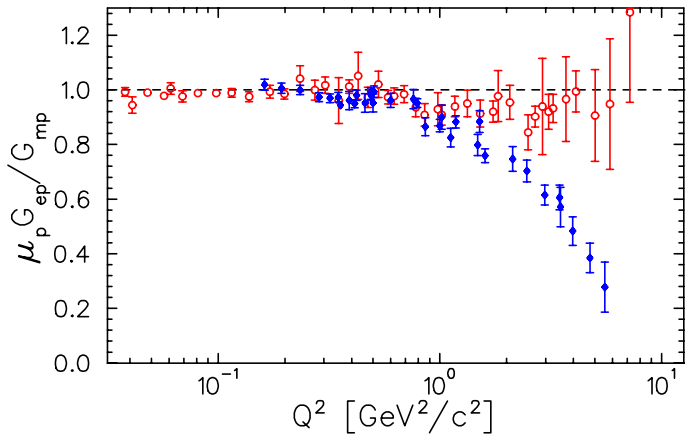}
\caption{\label{figboth}
Proton electromagnetic form factor data \cite{arringtonpc} for 
$\mu_{\rm p} G_{\rm ep} / G_{\rm mp}$
as obtained via the Rosenbluth separation (red circles) and via the 
polarisation transfer technique (blue dots).
The Rosenbluth data indicates scaling (see Eq.~(\ref{scaling1}))
whereas the polarisation transfer data can be fitted linearly according to
Eq.~(\ref{fit}).
}
\end{figure}

According to the polarization transfer measurements the ratio 
$\mu_{\rm p}G_{\rm ep}/G_{\rm mp}$ decreases approximately linearly with 
$Q^2$, reaching a value of 0.2 at $Q^2=6\,{\rm GeV}^2$, in contrast to the 
scaling behaviour (\ref{scaling1}) seen in Rosenbluth measurements.
A linear fit to the polarization transfer data yields \cite{recpol2}
\begin{eqnarray}
\label{fit}
\mu_{\rm p}G_{\rm ep} / G_{\rm mp} 
= 1-0.13(Q^2-0.04) \, ,
\end{eqnarray}
shown in Fig.~\ref{figboth}, together with the Rosenbluth data.
As a consequence of this discrepancy \cite{recpol,recpol2}
an improved Rosenbluth measurement (called 'SuperRosenbluth')
was carried out at {\sc tjnaf} \cite{superrosenbluth,superrosenbluth2}.
In order to reduce the effects of radiative corrections and other systematic
uncertainties due to beam fluctuations, the SuperRosenbluth experiment 
measured the H$(e,p)$ cross section.
In addition, the world Rosenbluth data was re-analyzed \cite{arrington}
as well as the world polarization transfer data, but neither data 
set revealed internal inconsistencies.\\

The apparent discrepancy between the proton form factor measurements from
Rosenbluth separation technique and from polarization transfer experiments
has led to different possible explanations.
Most approaches aim to explain the discrepancy in terms of the 
contribution from the two-photon exchange
{\sc tpe} diagrams (sometimes also called 'box diagrams', see Fig.~\ref{fig1})
\cite{blunden03,guichon03,blunden05} which can only be included
into the cross section approximately due to inelasticities on the proton
side of the two {\sc tpe} diagrams 
\cite{arrington,motsai,calan,maximontjon,blunden03,guichon03}.
Blunden {\em et al.}~can reduce the apparent discrepancy 
in $G_{\rm ep} / G_{\rm mp}$
by roughly a factor of 2 when including the proton ground-state only, with the 
Rosenbluth data moving down towards the polarization data \cite{blunden03}. 
Further hadronic calculations of the {\sc tpe} contribution, involving
more intermediate states ({\em e.g.}~the $\Delta$ resonance), are model 
dependent and only valid for small and intermediate values of $Q^2$, 
since for larger values of the momentum transfer more and more intermediate 
states have to be taken into account \cite{kondratyuk}.
Calculations involving {\em all} intermediate states 
can be carried out using generalized parton distributions,
relating them to virtual Compton processes on the nucleon.
They are valid for values of $Q^2\approx 1 \mbox{GeV}^2$ and larger when
a virtual photon starts to resolve point-like partons
\cite{chen,afanasev5,guidal}.
\\

The Rosenbluth technique is very sensitive to corrections 
depending on $\varepsilon$ and the {\sc tpe} is such a correction.
While the effects from {\sc tpe} are merely at the level of around
$5\%$, the contribution of $G_{\rm ep}$ is also just a few percent at
high $Q^2$, rendering the impact on  $G_{\rm ep}$ much larger.
So the effects from {\sc tpe} correction are magnified considerably
by the fact that the slope of the reduced cross section 
$\sigma_{\rm red}$ as a function of $\varepsilon$ is so small.
While the {\sc tpe} diagrams can usually be neglected in the cross section, 
they render the Born approximation invalid in the case of the Rosenbluth 
technique \cite{blunden03}, aiming at small values of $G_{\rm ep}$.\\

The non-negligible {\sc tpe} contribution to the reduced cross section
(\ref{redcrosslinear}) destroys its linearity towards small values of
$\varepsilon$ \cite{blunden03,kondratyuk,kondratyuk2}.
But no indications for such non-linearities have been found so far in Rosenbluth
measurements. 
Refs.~\cite{tomasi,tvaskis} set limits on the non-linearities.
Ref.~\cite{tvaskis} does not rule out the non-linearities predicted by some
calculations.
Howere, these tests do not constrain the linear part of the correction which
can modify $G_{\rm ep} / G_{\rm mp}$.
A very clean experimental access to the {\sc tpe} contribution would be provided 
by positron scattering.
But suitable positron beams with the necessary luminosities are not yet available.\\

The {\sc tpe} effect on the Rosenbluth data only providing a partial resolution of
the discrepancy raises the question: which other corrections to
the reduced cross section exhibit an $\varepsilon$-dependence leading
to a sizable effect?\\

In this letter we study radiative corrections to $e$-$p$ scattering as a 
possible source for the discrepancy discribed above.
While these radiative corrections are usually approximated 
we here apply an improved radiative correction procedure to
Rosenbluth data in order to evaluate the effect of the approximations 
on the discrepancy.
In Sec.~\ref{sec2} we introduce radiative corrections to $e$-$p$ scattering,
highlighting the most common approximations used in radiative correction
procedures.
In Sec.~\ref{sec3} we sketch an improved correction procedure which
partially removes these approximations.
In Sec.~\ref{sec4} we apply the improved radiative corrections to 
Rosenbluth data, showing that the approximations usually made in the 
treatment of radiative corrections to $e$-$p$ scattering data have 
little effect on the proton electric form factor as measured in Rosenbluth 
type experiments. \\

Full calculations of radiative corrections to order $\alpha^1$ have already been
calculated for radiation originating from the incident and the outgoing electron
\cite{afanasev100,afanasev200,afanasev300}.  
But the improved radiative corrections shown here go beyond order $\alpha^1$. \\

In accordance with Refs.~\cite{afanasev100,afanasev200,afanasev300}
we find that the improved corrections are small and do not contribute
significantly. 
Even though this is a negative outcome it 
provides an important clarification, as it has been discussed as a possible
explanation for the discrepancy between the two approaches. 
In addition it was found that the improvement is important in other observables. \\

\section{Radiative corrections to $e$-$p$ scattering}
\label{sec2}
For practical purposes radiative corrections (see Fig.~\ref{fig1}) 
to $e$-$p$ scattering data are usually carried out using approximations
\cite{schwinger,tsai,motsai,makins}. 
While hadronic contributions to the radiative corrections can still be 
included to a good accuracy, most procedures employ two approximations,
the soft-photon approximation ({\sc spa}) and the peaking approximation
({\sc pa}) in order to simplify the calculations 
\cite{makins,schiff,maximon}.\\

The {\sc spa} assumes that the emitted brems\-strah\-lung photon has no effect
on the hard scattering; this is justified (only)
in the limit where the brems\-strah\-lung photon
has vanishing energy $\omega^0$.
Consequently, in {\sc spa}, the cross section for emitting one soft 
brems\-strah\-lung photon factorizes into the elastic first-order Born
cross section ${\cal M}_{\rm Born}^{(1)}$, times the probability for 
emitting a soft brems\-strah\-lung photon \cite{landau,weinberg,itzykson}.
This factorization also applies in the case of multi-photon brems\-strah\-lung
\cite{gupta} where it translates into an exponentiation of the soft-photon 
contribution to the cross section \cite{landau,weinberg,yennie,makins}.
The exponentiation of the soft-photon contribution
renders radiative correction procedures much more straightforward, 
considerably simplyfing data analyses \cite{makins,weissbach1,weissbach2}.
However, in practice the {\sc spa} is applied to scattering
events accompanied by the multiple emission of photons with energies
which cannot be considered as 'soft' photons any more \cite{weissbach2}.\\

The {\sc pa} is based on the {\sc spa}. It further simplifies 
radiative correction procedures by assuming that the momenta of all brems\-strah\-lung 
photons are aligned with the emitting particles; it was first introduced 
by Schiff in 1952 \cite{schiff} for $(e,e')$ experiments. 
Later is was extended to inclusive $(e,e'p)$ scattering by Ent 
{\em et al.}~\cite{makins}.
The {\sc pa} is inspired by the observation that H$(e,e'p)$ data indeed 
show that the brems\-strah\-lung photons are emitted mostly along the 
directions of the incident electron ($e$), the scattered electron ($e'$), 
and the recoiling proton ($p$) \cite{makins,weissbach1}.
But the {\sc pa} overestimates the amplitudes of the photon peaks and cannot appropriately 
treat the kinematics and the evaluation of the form factors for those 
brems\-strah\-lung photons which deviate from the $e$-, $e'$, and
$p$-directions \cite{weissbach1,weissbach2}.\\

While {\sc spa} and {\sc pa} both exhibit shortcomings, together
these two approximations considerably 
simplify the numerical treatment of the radiative corrections
to $e$-$p$ scattering data.
For many purposes the two approximations are of good quality and
may be used without harm; but there are also experimental settings for which
the approximative application of radiative corrections do lead to inaccuracies
\cite{makins,weissbach1,weissbach2}.

\section{Improved radiative corrections}
\label{sec3}
As mentioned above the {\sc spa} considerably simplifies the treatment of 
multi-photon brems\-strah\-lung.
In fact an exact treatment of multi-photon brems\-strah\-lung 
without {\sc spa} is not feasible since higher-order brems\-strah\-lung diagrams 
cannot be included into radiative correction calculations to arbitrary order
(in $\alpha$).
It has, however, been shown that the {\sc spa} can partially be removed
from multi-photon radiative correction procedures by treating one 'hard' 
brems\-strah\-lung photon exactly while calculating the remaining photons
in {\sc spa} \cite{weissbach2}.
Because this novel approach combines brems\-strah\-lung photons treated
exactly with a kind of brems\-strah\-lung 'background' which is treated in
{\sc spa}, it is here referred to as the {\em combined} calculation.
It has been shown that this combined calculation is invariant under 
different methods of selecting
the 'hard' photon, which is treated exactly,
from a given multi-photon event \cite{weissbach2}.\\

\begin{figure}%[t]
\centering
\includegraphics[width=7cm]{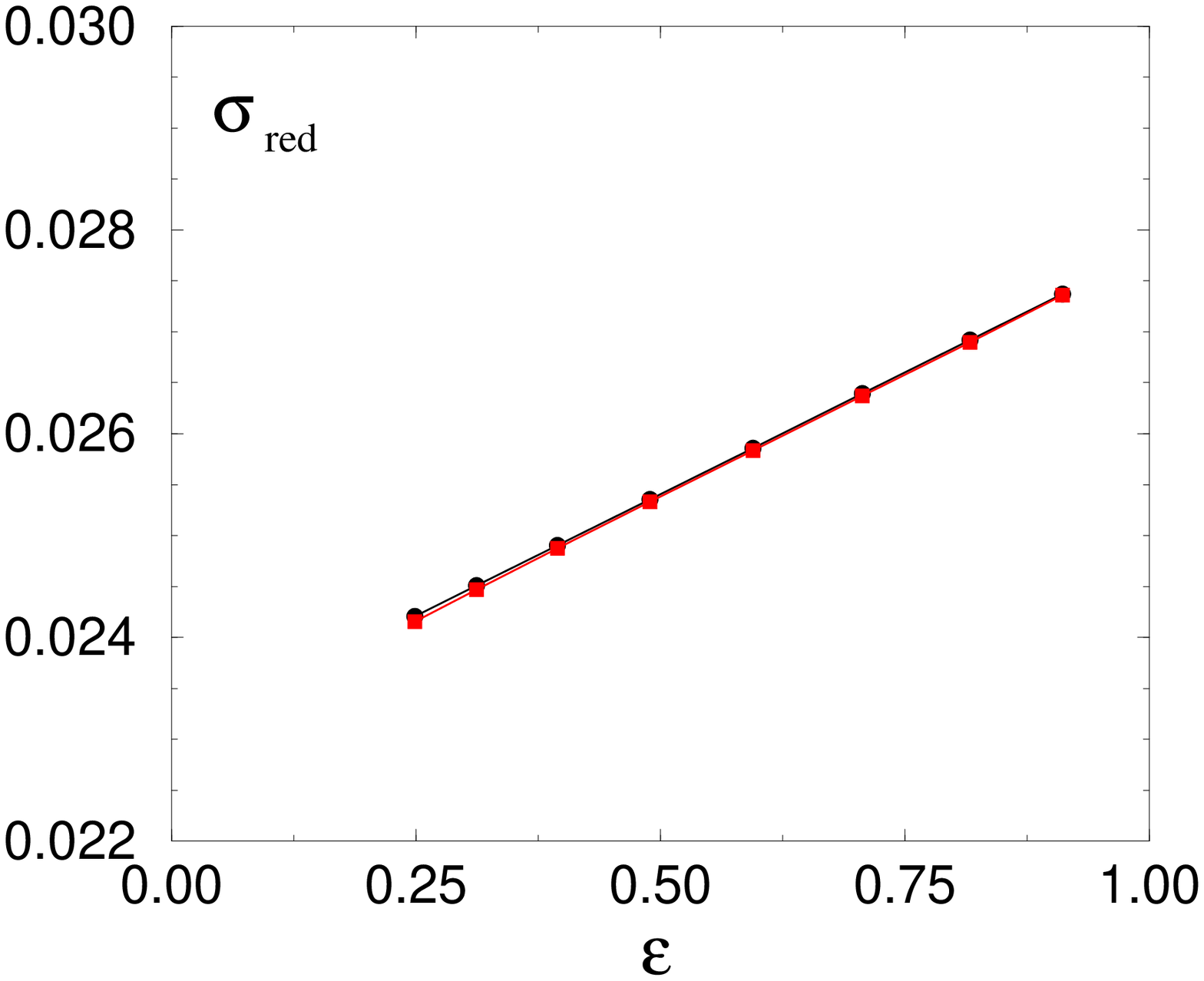}
\includegraphics[width=7cm]{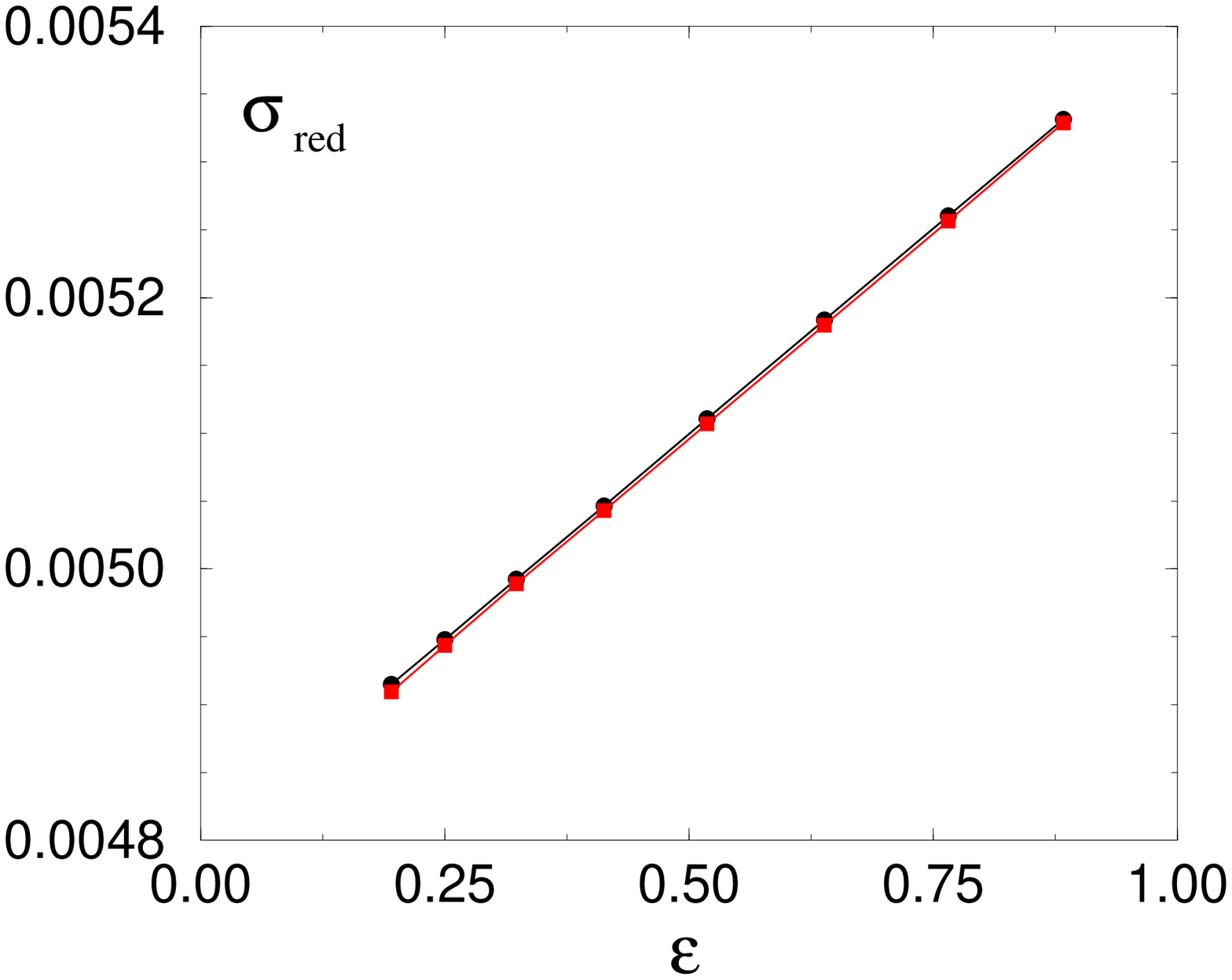}
\includegraphics[width=7cm]{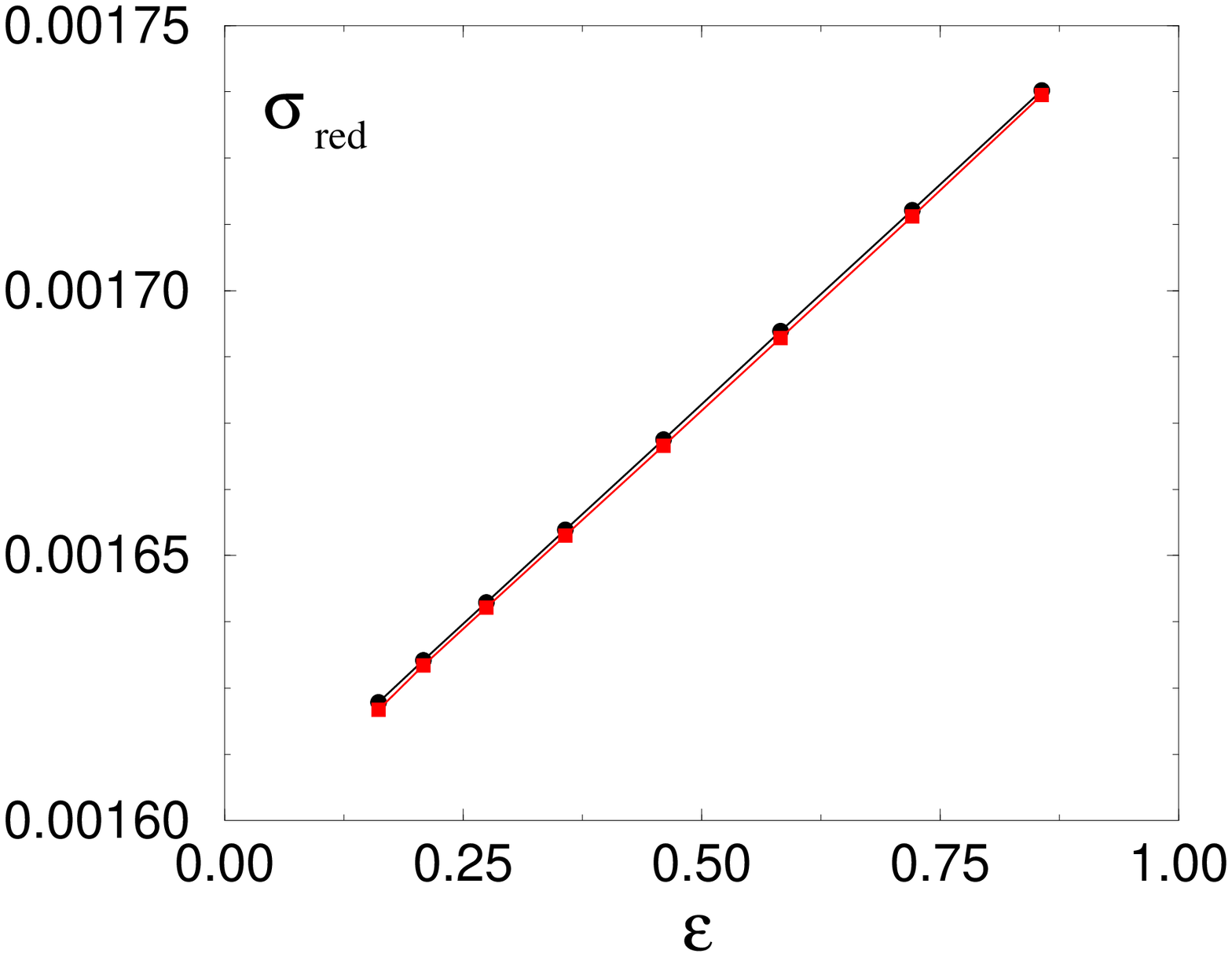}
\caption{\label{QQ2}
Rosenbluth plots: reduced cross section as a function of $\varepsilon$
at three different values of the momentum transfer ($Q^2=2.0\,{\rm GeV}^2$, 
$Q^2=4.0\,{\rm GeV}^2$, and $Q^2=6.0\,{\rm GeV}^2$).
The red lines (squares) show the reduced cross sections calculated using
the improved radiative corrections.
The black lines (circles) are hardly distinguishable from the red lines. 
They depict the reduced cross sections calculated using {\sc spa} and {\sc pa}.
One can clearly see that the improved radiative corrections have no
sizable impact on the Rosenbluth plots and hence on $G_{\rm ep}$.}
\end{figure}

It has further been shown that the {\sc pa} can fully be removed from
$e$-$p$ data analyses without large computational expense by
introducing a full angular treatment of the brems\-strah\-lung photons
\cite{weissbach1}.
The full angular treatment renders the assumption,
that all brems\-strah\-lung photons are either emitted in $e$-, $e'$-,
or $p$-direction, unnecessary.
Removing the {\sc pa} leads to improved kinematical treatment and to a more
systematic evaluation of the form factor \cite{weissbach1}.\\

Both improvements -- the partial removal of the {\sc spa} and the 
complete removal of the {\sc pa} -- can be done simultaneously and
have simultaneously been applied to $(e,e'p)$ data \cite{weissbach2}.
Together they are here referred to as the {\em improved} radiative
correction treatment.
The improved radiative correction treatment reproduces experimental ($e,e'p$) data 
more accurately than correction procedures fully relying on {\sc spa} and 
{\sc pa} \cite{weissbach2} and the question arises whether this effect can 
also be seen in Rosenbluth experiments.

\section{Results and discussion}
\label{sec4}
In order to recompute Rosenbluth plots using the improved radiative 
corrections we used an empirical fit to
the world Rosenbluth data \cite{bosted} and generated the reduced cross section
(\ref{defredcross}) with $Q^2=2.0\,{\rm GeV}^2$, with $Q^2=4.0\,{\rm GeV}^2$,
and with $Q^2=6.0\,{\rm GeV}^2$.
The comparison between existing $(e,e'p)$ Rosenbluth results with
a calculation based on the improved approach described here was done by
multiplying the reduced cross section (\ref{defredcross})
with a correction factor accounting for the differences between
the two radiative correction treatments.\\

Our calculations of the Rosenbluth results using standard radiative corrections 
(with full {\sc spa} and {\sc pa}) and of the improved approach was done separately for 
each momentum transfer squared. 
A Monte Carlo generator was used to sample multiple 
brems\-strah\-lung photons per scattering event (see also \cite{weissbach2}).
Using the techniques and assumptions described above the four-momenta of the 
scattered electron and the recoiling proton were calculated subsequent to the
emission of brems\-strah\-lung photons. 
Geometrically the detectors were treated as rectangular windows.
And the detectors were given a momentum acceptance.
Once an electron (proton) passed through the "window" of the electron (proton) detector 
its three-momentum was computed in order to check whether it was within the detector's 
acceptance range. Events with electrons (protons) outside the acceptance were given zero weights.
Weights of events with  electrons (protons) not passing through the "windows" were also set to 
zero.\\

In order to compare the two approaches the missing energy
was considered, binning the two event weights (one coming from the approximate
procedure, the other calculated via the improved procedure) in the vicinity 
of the total missing energy for each event.
\begin{table}
\begin{center}
\begin{tabular}{llll}\hline
$Q^2$      & $(G_{\rm ep}^{\rm approx.})^2$  & $(G_{\rm ep}^{\rm impr.})^2$ & deviation\\ \hline
$2.0\,{\rm GeV^2}$ & $4.772\times 10^{-3}$ & $4.816\times 10^{-3}$   & $+0.92\%$ \\
$4.0\,{\rm GeV^2}$ & $6.055\times 10^{-4}$ & $6.081\times 10^{-4}$   & $+0.43\%$ \\
$6.0\,{\rm GeV^2}$ & $1.660\times 10^{-4}$ & $1.662\times 10^{-4}$   & $+0.12\%$ \\ \hline
\end{tabular}
\end{center}
\caption{\label{tab3} 
Impact of the improved radiative corrections on the proton electric form 
factor.
The experimental errors on $G_{\rm ep}^2$ are usually large such that the 
deviations shown in this table are entirely negligible.}
\end{table}
As $(e,e'p)$ Rosenbluth experiments only consider the elastic
peak of the missing energy $E_{\rm m}$ up to energies of the
order of $20$ to $50\,{\rm Mev}$,
we integrated the two missing-energy distributions, obtaining the two total 
cross sections
\begin{eqnarray}
\label{corfac1}
\sigma_{\rm tot}^{\rm impr.}(E_{\rm m}\le 50\,{\rm MeV}) \, ,
\end{eqnarray}
and 
\begin{eqnarray}
\label{corfac2}
\sigma_{\rm tot}^{\rm approx.}(E_{\rm m}\le 50\,{\rm MeV}) \, .
\end{eqnarray}
These cross sections were used to correct the standard
reduced cross section (\ref{defredcross}) by multiplication with the
correction factor
\begin{eqnarray}
\label{corfac3}
\sigma_{\rm red}^{\rm ex}(\varepsilon)=
\frac
{\sigma_{\rm tot}^{\rm impr.}(E_{\rm m}\le 50\,{\rm MeV})}
{\sigma_{\rm tot}^{\rm approx.}(E_{\rm m}\le 50\,{\rm MeV})}
\sigma_{\rm red}(\varepsilon) \, .
\end{eqnarray}

The results are presented in Fig.~\ref{QQ2} and in Tab.~\ref{tab3}.
As one can see the correction factor (ratio on the r.h.s.~of Eq.~(\ref{corfac3}))
is very close to unity for the three kinematic settings considered here.
Given the large errors usually appearing in Rosenbluth measurements we 
can conclude here, that the improved radiative corrections have no visible 
impact on $e$-$p$ Rosenbluth data.\\

Improved radiative corrections to $(e,p)$ measurements
such as the SuperRosenbluth experiment are more difficult than
the corrections to $(e,e')$ and $(e,e'p)$ experiments since
the scattered electron's momentum (which is not measured) would have to 
be generated very efficiently in an additional Monte Carlo generator.
However, since the $(e,p)$ SuperRosenbluth data do not exibit any
deviations from other Rosenbluth measurements 
\cite{superrosenbluth,superrosenbluth2}, the improved radiative corrections 
should lead to similarly small (or even smaller) corrections as the ones 
shown here in Tab.~\ref{tab3}. \\

In conclusion, we observe that the approximate radiative correction
treatment (using {\sc spa} and {\sc pa}) does not have a sizable impact on
the reduced cross section as obtained via the Rosenbluth technique.
This may be due to the fact that the Rosenbluth technique considers a narrow range around 
the elastic scattering peaks where breams\-strah\-lung does not play the role it has in 
radiative tails.\\

\section{Outlook and further investigations}
\label{sec5}
Removing or partially removing approximations such as {\sc pa} and {\sc spa}
from radiative corrections in the way sketched here and elaborated further
in Refs.~\cite{weissbach1,weissbach2} could be carried over to radiative
corrections to initial-state radiation ({\sc isr}) experiments.
These collider mode experiments could yield independent measurements of
$G_{\rm ep}^2$ and $G_{\rm mp}^2$ by studying the
\begin{eqnarray}
\label{isrreaction}
e^+e^-\rightarrow \gamma p{\bar p} 
\end{eqnarray}
reaction, the photon being emitted by one of the leptons.
The {\sc isr} matrix element is smaller than the 
$e^+e^-\rightarrow p{\bar p}$ matrix element by a factor of 
${\cal O}(\alpha)$.
But 'meson factories' like {\sc da$\phi$ne}, {\sc cesr},
and {\sc kek-b} are pushing the luminosity frontier far enough 
to compensate for this suppression \cite{isr2,isr1}.
{\sc isr} experiments usually detect the hardest initial-state photon,
its angular distribution being found to be mostly aligned with the electron 
or the positron three-momentum \cite{aubert},
as the brems\-strah\-lung photons in $e$-$p$ scattering.
Softer initial-state photons constitute a background, as well as
final-state radiation ({\sc fsr}).
The latter can be suppressed by suitably chosen cuts.
{\sc fsr} seems to be negligible for $p{\bar p}$ production, except for the 
region close to threshold, where the Coulomb interaction becomes dominant 
\cite{isr2}.
Radiative corrections to $e^+e^-$ experiments, relying on {\sc pa} and 
{\sc spa}, have been carried to next-to-leading order using Monte Carlo 
simulations \cite{isr2,phokhara,schaelicke}.
As in the case of $e$-$p$ scattering, {\sc pa} and {\sc spa} used in
{\sc isr} experiments may be (partially) removable:
the {\sc pa} could be replaced by a full angular treatment \cite{weissbach1}
and the {\sc spa} could partially be removed by calculating one hard photon
from a given multi-photon event exactly;
and the remaining photons could be accounted for resorting to the {\sc spa}
\cite{weissbach2}.\\

\begin{acknowledgments}
The authors wish to express their gratitude to the 
{\sc Schweiz\-er\-ische Nationalfonds} which supported this work.
Moreover they are grateful to A.~Afanasev and J.~Arrington for
valuable discussions on the topics presented here. 
\end{acknowledgments}
\end{document}